\documentclass[useAMS]{mn2e}
\usepackage{times}
\usepackage{rotate}
\usepackage{graphics}
\usepackage{psfig}
\usepackage{lscape}
\input{epsf}
\newif\ifAMStwofonts
\AMStwofontstrue

\def\gsim{\mathrel{\hbox{\rlap{\hbox{\lower4pt\hbox{$\sim$}}}\hbox{$>$}}}}
\def\lsim{\mathrel{\hbox{\rlap{\hbox{\lower4pt\hbox{$\sim$}}}\hbox{$<$}}}}

\title[AGN space density at $z$=3]{The Space Density of Moderate Luminosity Active Galaxies at {\bf $z$}=3}
\author[K. Nandra et al.]{K. Nandra$^{1}$, E.S. Laird$^{1}$, C.C. Steidel$^{2}$ \\
$^1$Astrophysics Group, Imperial College London, Blackett Laboratory,
Prince Consort Road, London SW7 2AW, UK \\ 
$^2$California Institute of Technology, Pasadena, CA 91125, USA \\}
\date{}

\begin{document}

\maketitle
\label{firstpage}

\begin{abstract}
We present an estimate of the space density of Active Galactic Nuclei (AGN) at $z=3$. Combining deep X-ray data with Lyman Break Galaxy (LBG) colour-selection in the rest-frame UV makes for highly efficient identification of AGN in a narrow redshift range ($z\sim 2.5-3.5$). Using {\it Chandra} data from the Groth-Westphal Strip (GWS) and the Hubble Deep Field North (HDF-N), we find a total of 15 X-ray detected LBGs at $z\sim$3, the majority of which are unlikely to have been identified in blanket followup surveys of X-ray detected objects. We find the comoving space density of moderate luminosity AGN (MLAGN; $L_{\rm X}=10^{43-44.5}$ erg s$^{-1}$) at $z=3$ to be a factor $\sim 10$ higher than the most powerful objects. The available data are consistent with a roughly constant space density of MLAGN from $z=0.5-3$, and they are also consistent with a mild decline in the space density above $z=1$ as predicted by the luminosity-dependent density evolution models of Ueda et al. (2003). This strong AGN activity at $z=3$ argues against previous suggestions that the majority of black hole accretion occurs at low redshift. A further implication of our investigation is that, as far as can be determined from current data, the majority of the AGN population at $z\sim$3 are selected by the LBG dropout technique. Although this is sensitive to AGN with a UV-excess due to an accretion disk, it predominantly identifies star-forming galaxies. A significant fraction of X-ray sources seem likely to be hosted by these more typical LBGs, further strengthening the starburst-AGN connection at high redshift. 
\end{abstract}


\section{INTRODUCTION}
\label{Sec:Introduction}

\begin{table*}
\centering
\caption{X-ray detected Lyman Break Galaxies in the GWS. Details of the X-ray detected LBGs in the HDF-N can be found in Nandra et al. (2002) and Laird et al. (in preparation). 
Col.(1): {\it Chandra} object name;
Col.(2): {\it Chandra} catalogue ID (Paper I);
Col.(3): LBG survey name (Westphal-);
Col.(4): Redshift;
Col.(5): {$\cal R$} magnitude;
Col.(6): Positional offset in arcsec after applying astrometric shift (see text);
Col.(7): Flux (0.5-2 keV) in units of $10^{-15}$ erg cm$^{-2}$ s$^{-1}$, with error bars calculated based
on the Poisson error in the detected counts (see Paper I);
Col.(8): Luminosity (2-10 keV) in units of $10^{44}$~erg s$^{-1}$;
Col.(9): Hardness ratio;
Col.(10): Optical Classification
\label{tab:data}}
\begin{center}
\begin{tabular}{cccccccccc}
\hline
CXO GSS & X-ray & LBG &  z & {$\cal R$} & Off & Flux & $L_{\rm X}$ & HR & Class \\
 (2000) & ID & Name & & (AB) &  (") & (0.5-2 keV) & 2-10 keV & & \\
(1) & (2) & (3) & (4) & (5) & (6)  & (7) & (8) & (9) \\
\hline
J141747.4+523510 & c82& MD106 & 2.754 & 22.64 & 0.82 & $1.80^{+0.24}_{-0.21}$ & $1.41^{+0.18}_{-0.14}$ & -0.42 & QSO \\
J141755.2+523532 & c99 & D54 & 3.199 & 22.77 & 1.18 & $0.79^{+0.15}_{-0.13}$ & $0.88^{+0.17}_{-0.14}$ & -0.28 & QSO  \\
J141757.4+523106 & c104 & M47 & 3.026 & 24.30 & 0.75 & $0.41^{+0.12}_{-0.10}$ & $0.40^{+0.12}_{-0.10}$ & -0.10 & AGN \\
J141800.9+522325 & ---& M10 & --- & 25.31 & 0.96 & $0.25^{+0.08}_{-0.06}$ & $0.24^{+0.08}_{-0.06}$ & -1.00 & --- \\
J141801.1+522941 & c113 & oMD13 & 2.914 & 23.33 & 0.80 & $0.44^{+0.13}_{-0.10}$ & $0.35^{+0.12}_{-0.09}$ & -1.00 & QSO \\
J141811.2+523011 & c128 & C50 &  2.910 & 23.96 & 0.33 & $0.25^{+0.10}_{-0.07}$ & $0.20^{+0.09}_{-0.07}$ & -0.16 & GAL \\
\hline
\end{tabular}
\end{center}
\end{table*}

Much evidence now points toward a strong connection between the formation and evolution of galaxies and their central black holes. If major episodes of star formation in galaxies are triggered by mergers (e.g. Sanders \& Mirabel 1996), there is the prospect that this will make gas available in the galactic nucleus where it can feed a black hole. The end products of these merger-driven starburst events -- galaxy bulges - host dormant black holes today, and this process can therefore plausibly result in the observed relation between black hole mass and host galaxy bulge velocity dispersion (Ferrarese \& Merritt 2000; Gebhardt et al. 2000). An important consequence of this is that AGN activity should evolve concurrently with merger-driven star formation and galaxy building. Evolutionary studies show broad agreement with this idea, with the optical/soft X-ray QSO space density and the global star-formation density having a similar form (e.g. Boyle \& Terlevich 1998).

There may be differences in detail, however. In particular most studies have indicated that the QSO epoch ended earlier (at $z \sim 1.5-2$) than the decline of vigorous star formation ($z \sim 1$; Lilly et al. 1995; Madau et al. 1996). Star formation is still going strong at $z=4$ (Steidel et al. 1999; hereafter S99), but optical QSO activity appears to drop off at $z>3$ (e.g. Schmidt, Schneider \& Gunn 1995). On the other hand, ROSAT soft X-ray surveys show no clear decline in X-ray selected QSOs at high redshift (Miyaji, Hasinger \& Schmidt 2000). This last disagreement hints that the results on AGN evolution are sensitive to the selection and analysis methods. It should also be borne in mind that they are based only on powerful QSOs. We now know (e.g. Steidel et al. 2002 hereafter S02; Cowie et al. 2003 hereafter C03)  that the majority of nuclear activity occurs in more moderate luminosity AGN (MLAGN; $L_{\rm X} \sim 10^{43}$ erg s$^{-1}$). Furthermore, previous results regarding AGN evolution now have to be re-examined in the light of new data from ultra-sensitive X-ray surveys.

The deepest {\it Chandra} surveys provide a highly efficient way of detecting MLAGN out to very high redshift ($z>4$, e.g. Barger et al. 2003b). They are also relatively unbiased, compared with optical or soft X-ray surveys, against obscured objects. The first attempts at determining AGN evolution from deep {\it Chandra} surveys (C03; Ueda et al. 2003 hereafter U03; Silverman et al. 2004; Barger et al. 2005) confirm a rapid rise in the number of high luminosity AGN ($L_{\rm X}>10^{44}$ erg s$^{-1}$) towards high redshift. This evolution is consistent with results from optical and soft X-ray surveys (e.g. Boyle, Shanks \& Peterson 1988, Boyle et al. 1993). 

The story for MLAGN is, however, apparently quite different. For example, C03, based on a sample of X-ray selected objects suggested that the majority of MLAGN activity occurs at relatively low redshift ($z<1$). Indeed their conclusion was that the majority of black hole accretion in galaxies occurs at these low redshifts. The total AGN space density presented by these authors shows an apparent decline above $z\sim 1$, although this becomes less certain when the effects of spectroscopic incompleteness are taken into account. These results are very similar to those presented by Silverman et al. (2004), in a shallower, but wider survey, showing a peak of MLAGN activity at $z\sim 1$ and a strong apparent decline at higher redshifts. Such a conclusion would be fundamental, and apparently at odds with the presumed linkage between black hole growth, galaxy formation and mergers. Milder high redshift evolution of MLAGN, but still with the same sense, was presented by U03. 

These X-ray evolution results have been based on the approach of taking X-ray selected samples, following them up with optical spectroscopy, and then attempting to correct for both the X-ray and optical incompleteness. Such an approach could be problematic if the incompleteness corrections are large, as they are very difficult to determine. There may furthermore by systematic biases due to the difficulty of optical identification sources in the redshift ``desert'' from $z\sim 1.5-2.5$, where few strong spectral features are present and typical X-ray sources start to become very faint optically. Here we adopt a different approach to the determination of AGN evolution, by starting with an optically-selected sample of Lyman Break Galaxies (LBGs; Steidel et al. 2003 hereafter S03) and determining their X-ray properties. 

\section{X-ray detected Lyman Break Galaxies}

Analysis of the the {\it Chandra} deep X-ray survey of the GWS has been presented by Nandra et al. (2005; hereafter paper I). Briefly, the observation was of $\sim 200$ks exposure with the ACIS-I instrument ($17^{\prime} \times 17^{\prime}$). Standard screening criteria were applied to the data, which were analysed in the 0.5-7 keV range. The X-ray point source catalogue in Paper I comprises 158 sources down to a limiting flux of $1.1 \times 10^{-16}$ erg cm$^{-2}$ s$^{-1}$ in the 0.5-2 keV band, corresponding to a limiting luminosity of approximately $10^{43}$~erg s$^{-1}$ in the rest frame 2-10 keV band at $z=3$. Throughout this paper we adopt a flat cosmology with $h=0.7$ and $\Omega_{\rm \Lambda}=0.7$.

A Lyman Break survey has also been performed in this same area of the GWS, described by S03 (referred to as the ``Westphal" field by those authors). There are a total of 335 Lyman Break Galaxies identified by colour selection in this field, of which 188 have confirmed spectroscopic identifications as $z\sim 3$ galaxies. The interloper fraction (mostly stars) in these LBG surveys using the Steidel et al. filter set is, however, very small indeed ($\sim 4$~\%; S03), and we therefore consider the entire photometric LBG sample in our analysis, rather than the spectroscopic subset. 

Correlating the X-ray catalogue in Paper I with the LBGs of S03, after accounting for an astrometric offset (+1.15" in RA and +1.71" in DEC), results in 5 matches within a radius of 1.5". The catalogue in Paper I contains all sources down to a probability threshold of $4\times10^{-6}$, at which level approximately 1 false source is expected in the entire catalogue. As we are only interested in the X-ray emitting LBGs, however, a lower threshold can be used without fear of introducing spurious sources. Cross-correlation of the LBG catalogue with a lower-threshold catalogue ($2 \times 10^{-4}$) yields an additional LBG, MM10, associated with an X-ray source with false probability $4.7 \times 10^{-6}$, in fact only just above the threshold limit of Paper I. Considering there were only 335 test positions and a small search radius the probability that the X-ray detection is spurious is very small indeed. The probability of finding an X-ray source at a threshold $<2 \times 10^{-4}$ at the position of a known LBG at random is $1.3 \times 10^{-3}$, so we would expect $\sim 0.4$ chance matches. Details of all 6 X-ray detected LBGs in the GWS are shown in Table~\ref{tab:data}.  \footnote{In passing, we note that Westphal-MD11, an unusual, submm luminous LBG (Chapman et al. 2002), is not detected in our X-ray image, meaning if it does contain an active nucleus it must either be of relatively low luminosity ($<10^{43}$ erg s$^{-1}$), or Compton thick (e.g. Reynolds et al. 1994; Matt et al. 1996).}

We have combined the GWS X-ray LBG data with those from the Hubble Deep Field North (HDF-N). 
Nandra et al. (2002) reported the detection of 4 LBGs in the HDF-N based on the 1Ms {\it Chandra} observation. Deeper (2Ms) X-ray data for this field have now been presented by Alexander et al. (2003; hereafter A03). Here we use our own independent analysis of the 2Ms data, in which a total of 9 LBGs have been detected in the S03 survey area. The reader is referred to Laird et al (2005) for further details. Our final sample of X-ray detected LBGs in the two fields consists of 15 objects. Of these 15 objects, 12 have spectroscopic identifications. The optical types are equally split between broad-line objects (i.e. QSOs), narrow-line AGN, and ``normal'' LBGs according to the classifications of S03. 

\section{The comoving space density of AGN at $z$=3}

\begin{figure*}
\begin{center}
\rotatebox{270}
{\scalebox{0.6}
{\includegraphics{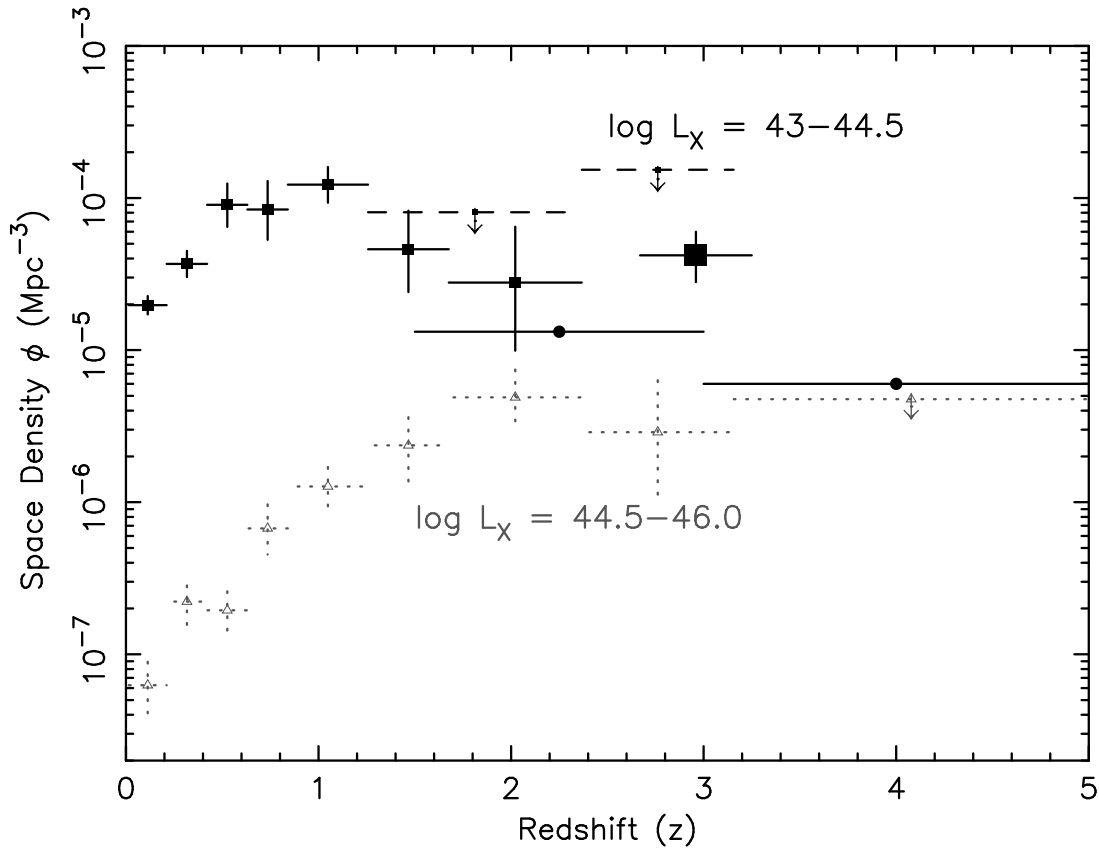}
}}
\caption{Comoving space density of AGN from Ueda et al. (2003), and our own LBG X-ray observations. High luminosity AGN (open triangles, dotted error bars) show strong positive evolution to high z, with a ``peak'' of the QSO epoch begining at $z=1.5-2.0$. Moderate luminosity AGN (squares), are more numerous at low z.  Our LBG point (large square) shows that MLAGN activity is still extremely strong at $z=3$, with a $\sim$constant space density at high redshift, similar to the evolution of the high luminosity objects.  Our data point is consistent with both the the upper limit to the high redshift space density from U03 (dashed lines, arrows) and their luminosity-dependent density evolution model. Also shown are estimates of the space density determined by integrating the luminosity functions at $z=1.5-3$ and $z=3-5$ presented by Barger et al. (2005; circles) in the MLAGN range. We find no evidence for the strong decline in the space density of MLAGN above $z \sim 1$ claimed by C03 and Barger et al. (2005). 
 \label{fig:sdens}}
\end{center}
\end{figure*}

To transform our observations of X-ray detected LBGs into an estimate of the AGN space density we need to determine the cosmological volume sampled by the survey. Detailed estimates of the completeness-corrected survey volumes of the LBG surveys have already been calculated and presented by S99, who used them to estimate the global star-formation rate density. As we consider only AGN hosted by LBGs it should be reasonable to approximate the volume of our combined LBG/X-ray survey with these values. The X-rays are simply used to tell us whether the LBGs host AGN or not, and to determine their X-ray luminosity. We must, however, consider the completeness of the X-ray detections when estimating the total volume. The calculation of sensitivity versus area has already been described in Paper I, and we use those results here. The total space density is given by $$\phi(z) =  \sum^{N}_{i=1}1/A_{i}(F_{\rm X}) V_{\rm i} (\cal{R})$$ where $A_{\rm i}$ is the summed area in arcmin of the GWS and HDF surveys sensitive to source $i$, with 0.5-2 keV (observed frame) flux of $F_{\rm X}$, and $V_{\rm i}$ is the effective survey volume per square arcmin given in Table 3 of S99, which is a function of the object's magnitude in the $\cal{R}$ band. The summation is over the number of objects, $N$, in a particular X-ray luminosity range. 

Our data are most suitable for determining the space density of moderate luminosity AGN, with $\log L_{\rm X} = 43-44.5$~erg s$^{-1}$ in the 2-10 keV band. This luminosity range also matches the medium luminosity band presented by U03, which means we can make a comparison with their results. More luminous objects are too rare to be detected in significant numbers in our limited survey area. Indeed no LBG with luminosity $>10^{44.5}$~erg s$^{-1}$ is detected in either the GWS or HDF-N. The area of the survey sensitive to objects fainter than $10^{43}$~erg  s$^{-1}$ is small, comprising the central region of the HDF-N only. All 6 of the LBGs in the GWS and 4 of the 9 LBGs in the HDF-N are in the MLAGN range. The MLAGN in the HDF-N comprise the 4 detections in the 1Ms HDF-N {\it Chandra} data (Brandt et al. 2001) presented by N02, with the 5 new detections in the 2Ms data having $42< \log L_{\rm X}<43$ (Laird et al. 2005).

We find the space density of the moderate luminosity AGN ($\log L_{\rm X} = 43-44.5$ erg s$^{-1}$) to be $\phi = 4.2^{+1.8}_{-1.4}  \times 10^{-5}$ Mpc$^{-3}$, where the errors purely reflect the Poisson uncertainty considering there are only 10 objects in the bin, and are 68 per cent confidence limits calculated according to the prescription of Gehrels (1986). We note that the correction for X-ray incompleteness is quite small ($\sim$20~\%) due to the high sensitivity of the GWS and HDF-N surveys. Fig.~\ref{fig:sdens} shows this estimate together with the corresponding data from U03. We find that MLAGN activity is still very strong at $z=3$. Indeed, while the data points at $z=0.5-1$ are higher than those from $z=1-3$, the data are formally consistent with a constant space density from $z=0.5-3$.  Fig.~\ref{fig:sdens} shows that our data are consistent with the upper limit from U03 at the same redshift, and they are also consistent with their model, which shows a mild decline in the MLAGN space density above 
$z\sim 1$. 

The other previous estimate of the space density of MLAGN at these redshifts has been presented by C03. They consider the evolution of the number  density of AGN with $L_{\rm X}>10^{42}$~erg s$^{-1}$ (2-8 keV) out to $z\sim 3.5$. All 15 of our LBGs are above this luminosity limit, and applying the volume and completeness calculations gives a space density estimate $\phi = 1.07^{+0.36}_{-0.27}  \times 10^{-4}$ Mpc$^{-3}$, more than double the $\phi$ from MLAGN only. This is approximately an order of magnitude higher than the C03 estimate and roughly equal to their upper limit. This work has recently been updated by Barger et al. (2005), who have presented luminosity functions in the $z=1.5-3$ and $z=3-5$ ranges, incorporating both spectroscopic and photometric redshifts. We have provided a comparison with those results in Fig.~\ref{fig:sdens}, by making an approximate fit to and integrating their luminosity functions over the  $\log L_{\rm X} = 43-44.5$ erg s$^{-1}$ range. After accounting for the differences in adopted cosmology in X-ray bandpass, our $z\sim 3$ result is a factor $\sim 3$ higher than that found the Barger et al. (2005) in the $z=1.5-3$ range, and $\sim 7$ times higher than their $z=3-5$ point. We discuss the reasons for this discrepancy below. Comparing our $z=3$ estimate to the C03 and Barger et al. (2005) results at $z=1$ once again is consistent with, and supportive of, the hypothesis that AGN activity remained roughly constant between $z=1$ and $z=3$. 

\section{DISCUSSION}

We have shown, by combining deep X-ray data with the LBG surveys of S03, that the number density of MLAGN is roughly a constant function of redshift, above $z\sim 1$.  
The overall evolution of the AGN space density shows very good agreement with the evolution
of the star-formation rate density (e.g. Lilly et al. 1995; Madau et al. 1996; S99), as first suggested by Boyle \& Terlevich (1998), and supports the idea that the two processes have a common trigger, most probably in galaxy mergers. The MLAGN evolution is similar to that of higher luminosity AGN (i.e. powerful QSOs), except the decline in QSO activity begins at an earlier epoch ($z\sim 1.5-2$). This latter conclusion has been persuasively argued in previous evolutionary studies (C03, U03). Our results add to this by showing that AGN activity remains strong out to $z=3$, contrary to any indication that there might be a reduction in MLAGN activity at high redshift. The LBG data are, however, consistent with a mild decline in the space density towards high $z$, for example as predicted by the luminosity-dependent density evolution model of U03. 

It should be emphasized that our method of determining the AGN space density is completely different to that employed in the earlier studies. We are using a rest-frame UV--selected sample, over a narrow range of redshift, with the cosmological volume determined from the optical selection function. The X-ray data are used only to determine whether the LBGs host an AGN, and to calculate their luminosity for comparison with other surveys. We have also applied a correction for X-ray incompleteness, but this is relatively small. C03 and U03 take X-ray selected samples at all redshifts, and follow these up with optical spectroscopy, correcting for X-ray incompleteness only. Spectroscopic completeness is assessed by assigning any optically unidentified objects to a particular redshift bin, to estimate an upper limit. 

There are three major advantages of our method compared to blanket followup surveys. The first is that the LBG selection function is very well defined. As discussed by S99,  both the photometric incompleteness and, to a lesser extent the colour incompleteness can be assessed in the calculation of the effective survey volume. The second is that, as we are dealing with a small subset of objects in a specific redshift range, we can be more lax about our X-ray detection criteria. For example, 5 of the 9 X-ray detected LBGs in the HDF-N appear in the A03 catalogue. The detection threshold of this purely X-ray based catalogue was necessarily more conservative in order to avoid an unacceptable number of spurious X-ray sources. For the subsample of LBGs, however, a lower threshold can be applied making our X-ray detections more complete, which can be crucial if we are dealing with a population of sources which are close to the detection limit, as we indeed are with high-redshift MLAGN. Finally, using LBG colour selection largely circumvents any issues of spectroscopic completeness. Around $\sim 50$~per cent of the GWS/HDF-N LBG sample are spectroscopically confirmed, and the probability that the remaining photometrically identified LBGs are $z\sim3$ galaxies is very high ($\sim 96$~\%). Our redshift determinations should therefore be essentially complete to $\cal{R}$=25.5. Such limits are extremely difficult to reach in blanket followup surveys, even using 8m-class telescopes. 

Given the small redshift range, deep X-ray data and comprehensive optical coverage for our sample all corrections for completeness are relatively small. The same cannot necessarily be said for the blanket X-ray followup surveys and it seems most plausible that uncertainties in these corrections account for any discrepancy between these surveys and ours. For example, C03 present an estimate of the high redshift ($z=2-4$) evolution of AGN with $L_{\rm X}>10^{42}$ erg s$^{-1}$. Our result is about an order of magnitude above theirs. At high $z$, the C03 results rest entirely on the data in the 1Ms HDF-N, as the other fields used by C03 are insensitive to MLAGN at high redshift.  None of the 1Ms HDF-N survey area is, however, sensitive to $z=3$ AGN with $L_{\rm X}=10^{42}$~erg s$^{-1}$. As we have shown above, even the analysis of the 2Ms data by A03 is highly incomplete in the X-ray at low luminosity, missing 80\% (4/5) of LBGs in the $L_{\rm X} = 10^{42-43}$~erg s$^{-1}$ range. A comparison with the results of Barger et al. (2005), who used the 2Ms HDF-N and the 1Ms CDF-S combined to determine the high redshift X-ray luminosity function for MLAGN, confirms this. Our methodology minimizes incompleteness and bias in this very low luminosity range. It is not, however, immune to either effect, and in particular the very low count threshold we have employed leads to large uncertainties in the fluxes, and hence luminosities. 


The other way that high z AGN can be missed is due to incompleteness of the optical spectroscopic followup. Of the 5 LBGs detected in the X-ray catalogue of A03, only 2 have redshifts determined in the optical followup of Barger et al. (2003a). This effect has been accounted for by C03 and U03 by giving upper limits determined by assigning all the unidentified objects to a given redshift bin. These limits should be conservative and indeed are consistent with our space density value at a comparable redshift.  
Barger et al. (2005) added in photometric redshifts to their study, and still find fewer $z=3$ AGN than we do. Our LBG point shows that the many of the unidentified objects in these previous are likely to be in the highest redshift MLAGN bin. The comparison with C03 is even more striking, as we would have to assign all the unidentified objects to the $z\sim 3$ bin to maintain consistency with their results. Much of this may be accounted for by X-ray incompleteness, as discussed above, but it also seems very likely that the majority of unidentified objects in the followup surveys are indeed at high z. The fact that our number density at $z=3$ is so much higher than that of C03 also implies that the luminosity density at that redshift given by those authors has been underestimated. Simply scaling their upper limit gives a luminosity density in black hole accretion at $z\sim 3$ significantly higher than at $z\sim 1$. This apparently contradicts the conclusion of C03 that the majority of black hole accretion has occurred relatively recently ($z<1$). 

While our space density estimate differs from some previous work, a crude sanity check suggests it is of the correct order. Considering all magnitude ranges presented in S99, the space density of all LBGs can be estimated as $\phi(LBG)=1.2 \times 10^{-3}$ Mpc$^{-3}$. The optical AGN fraction in LBGs is 3\% (Steidel et al. 2002) and a similar number is obtained when considering the X-ray detection rate (Nandra et al. 2002). Combining these we arrive at a number density of AGN in LBGs of $4 \times 10^{-5}$ Mpc$^{-3}$, very similar to the estimate presented above. We do rely on the completeness-corrected volumes given by S99, which are not free from uncertainty. In particular, the selection function for both broad-line QSOs (Hunt et al. 2004) and narrow-line AGN (Steidel et al. 2002) is likely to be different to that from non-AGN LBGs. They would, however, have to be underestimated by almost an order of magnitude  to bring our number density into line with that of, e.g., C03. 



Our methodology has the disadvantage that it will miss any AGN which are too faint to be selected in the optical survey (the limit of which is $\cal{R}$=25.5) and/or that have colours which fail the LBG selection criterion. Indeed, as also pointed out by Hunt et al. (2004), there are 3 X-ray sources with $2.5<z<3.5$ reported by Barger et al. (2003a) which are not picked up in the LBG survey. This incompleteness should, however, be accounted for in the effective volume calculation. The maximum effective volume of the LBG survey (i.e. at bright magnitudes where the survey is highly complete) is 1370 Mpc$^{3}$ arcmin$^{-1}$, compared to a total volume from $z=2.5-3.5$ of  $3250$ Mpc$^{3}$ arcmin$^{-1}$. Thus, while we only expect the LBG selection function to pick up about 40\% of the objects at $z=2.5-3.5$ compared to a top hat function, this should be accounted for in our completeness-corrected volume. On the other hand, there may be AGN missed by the LBG selection for the more fundamental reason that they do not exhibit any significant flux in the rest-frame UV.  If this is a strong effect, however, it would result in an even {\it higher} space density than we estimate. 

If indeed there is a significant population of AGN at $z=3$ missed by the LBG selection they must also be missed in the X-ray surveys. There could plausibly be obscured AGN with very red optical colours, perhaps making up part of the bright submm-galaxy population, which has little overlap with LBGs (e.g. Chapman et al. 2000). If they are very heavily obscured, they may even be difficult to detect in the X-rays. It should be noted, however, that in the 2-7 keV {\it Chandra} band we sample a very hard bandpass, above $\sim 10$ keV in the rest frame. This should be largely immune to all but the heaviest, Compton thick obscuration. Indeed, even if the direct continuum is completely suppressed, $>50$~\% of the X-ray flux in the 10-30 keV band should emerge via Compton scattering (assuming a slab geometry slab with 2$\pi$ solid angle; George \& Fabian 1991).  Compton thick sources should therefore be detectable at $z=3$, and indeed several of our detected LBGs have X-ray colours consistent with a Compton reflection dominated spectrum.

The implication is that, at least at $z=3$, rest-frame UV colour selection on its own is an efficient way of selecting AGN, as has also been argued by Steidel et al. (2002). Obscured AGN or those with bright host galaxies drowning out any nuclear emission lines (e.g. Moran, Filippenko \& Chornock 2002) may be hard to identify as AGN in the spectroscopic followup, however. Indeed about 1/3 of objects in the LBG surveys have no other indication of AGN activity than their X-ray emission. Steidel et al. (2002) have furthermore argued that the properties of narrow-line AGN which host LBGs are consistent with those without clear nuclear activity. This, in combination with the space density results, argues that a large proportion of the AGN population at $z=3$ are hosted by normal LBGs. As the latter have high star formation rates, this again supports the idea that much AGN activity in galaxies at $z\sim 3$ occurs in galaxies with active star formation. On the other hand Hunt et al. (2004) have discussed the rest-frame UV properties of broad-line QSOs detected in the LBG surveys. These do not show any stellar absorption line signatures seen in typical LBGs and the nature of their  host galaxies are not currently known. 

Regardless of the optical spectra, our results show a large population of AGN selected in the LBG surveys. Indeed,  we have no evidence to suggest that there is a significant population of AGN which is not. There are also two LBGs which have AGN emission lines in the rest-frame UV which remain undetected in the 200ks GWS X-ray exposure. One such object was also undetected in the 1Ms HDF-N observation (Nandra et al. 2002), though it has now been detected in the 2Ms exposure. This should not be considered surprising, given that the X-ray-to-optical flux ratios of AGN cover at least two orders of magnitude. The clear message from the work described in this paper is that comprehensive deep survey  data in a both wavebands is required before a complete picture of nuclear activity in galaxies, or its evolution, can emerge. 

\section*{Acknowledgements}

We thank Y. Ueda, J. Silverman and P. Green for providing clarifications and information regarding their space density calculations, and Steve Warren and Christian Wolf for helpful discussions. ESL thanks PPARC for support in the form of a Research Studentship. We thank the referee, Dave Alexander, for helpful comments. 


\bsp

\label{lastpage}

\end{document}